# Voice Packet Performance Estimation through Step Network Using OPNET

**Taskeen Zaidi**,
Associate Professor,
Department of Computer Science,
SRM University (U.P.) 225003, India
E-Mail:taskeenzaid867@gmail.com

**Nitya Nand Dwivedi**,
Research Scholar,
Department of Computer Science,
SRM University (U.P.) 225003 India,
E-Mail: nityananddwivedi29@gmail.com

*Abstract-* VoIP transfer voice over networks such as LAN. This technology is growing rapidly due to support of existing network infrastructure at low cost. Various simulations have been done and it is observed that by increasing the VoIP client, packet length and traffic arrival rate the performance of step network affected. In the current work packet dropped, packet received, voice traffic sent and end-to-end delay is estimated for various queuing disciplines like PQ, FIFO and WFQ. It is depicted that queuing disciplines effects the applications performance and utilization of resources.

## I. INTRODUCTION

Queuing disciplines are implemented at the router. A good queuing discipline treats packet transfer in a fair and efficient manner that will be helpful to improve the quality of service. As FIFO is based on first in first out so it is quite similar to queue waiting at the bus stop for bus. PQ offers priority to real time packet transfer whereas WFQ offers weight and priority both to the packets.

## II RELATED WORK

A method for identification of recurring patterns in various platform independents environment is represented [1]. The performance of virtualized network is estimated automatically on various hardware with efficient resources utilization. The WSN algorithms are modeled and analyzed by using some techniques and then these techniques are applied to estimate the performance and checking Optimal Geographical Density Control (OGDC). The Monte Carlo simulation is further used to estimate OGDC. The simulation was done on 800 sensor nodes and the results were compared on NS2 simulation tool [2].

A brief overview of networks, networks architecture, topologies, protocols, models, layers, techniques, functions of layers are well described [3]. A brief explanation of the terms bandwidth, latency, throughput is described in [5]. The step topology model was validating by Zaidi and Saxena [6] by using object oriented model based on UML.A state diagram was converted to FSM and test cases were generating to validate step model for distributed computing environment. Structural topology optimization with benchmark studies was done [7]. The network topology models, structures, basic abstraction principle using graph theory, network topologies characterization and approaches for modeling the topology on internet was well explained by Wahlisch [8]. The performance of step topology [9] was measured by transferring large size audio and video packet in bytes through simulation tool. The performance of step topology was estimated on a simulation tool by Zaidi and Dwivedi [10] through various queuing disciplines implemented on routers and it was observed that for normal packet transfer over a network FIFO is best and for large packet size transmission and real time interactive packet PQ and WFQ is best. Transition mechanism was implemented and simulated by authors [11] according to organizational demand and a manual tunnel was created for secure data communication and it was depicted that tunnel provides better QoS to the ISP's, Data centres for packet transmission.[1]

## III. BACKGROUND
### A. Step Topology

A new topology called as step topology for static interconnection of handheld devices in steps as desktop computer systems,

**Figure 1. OPNET Representation of VoIP Step Network mode**

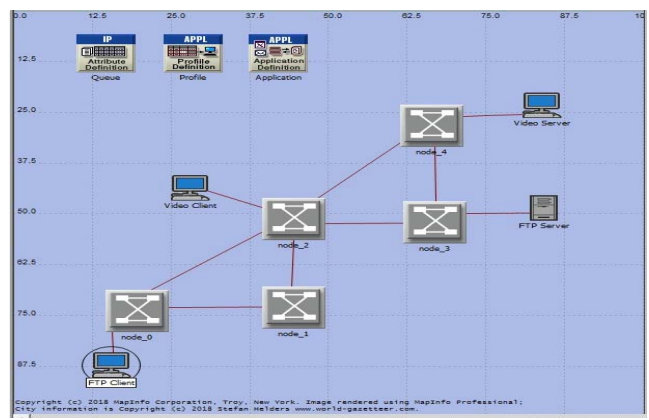

laptops, mobile, etc. This topology is modification of bus topology and by varying the steps we interconnect N numbers of computer systems. This topology works well if the link between two computers connected through bus topology fails. In this topology if individual node is busy then tasks can be executed on next node by using message passing technique in distributed manner and the devices can be connected in static as well as dynamically in adhoc manner. A view of step topology through OPNET simulation tool is shown in figure 1. The detailed overview about step topology can be found in [4].

### B. OPNET Modeler

Optimized network engineering (OPNET) tool is used to simulate the performance of the network . It has inbuilt models, protocols and devices that create and simulate different networks. It is a open software which has features like applications troubleshooting, protocol modeling, traffic modeling, validation of application, estimating performance of complex network systems. It has discrete event simulation workflow that helps in creation, import, and configuration of topologies as well as network traffic and is also published results with statistics report. Key features of OPNET are:

(i) Network model.
(ii) Process Modeling
(iii) Simulation kernel
(iv) Event based simulation.
(v) Link modeling.

To simulate performance of various scheduling techniques for VoIP, the architecture of step network is shown in Table 1.

**Table 1. List of Events for Transition Table**

| Application | ToS | Description |
|---|---|---|
| FTP and VoIP nodes | Best effort(0) | Connected to routers by 10 BaseT (10 Mbps) links. |
| VoIP application | Intercativevoice(6) | PCM Quality Speech" for voice, "Interactive Voice. |

### IV. Queuing Disciplines

For resources allocation management, routers in a network implements queuing disciplines so that packets waiting to be transmitted must buffered. The queuing discipline controls the packet transmission as well as packet dropped. It also effects the latency in a network. Some of the queuing disciplines are as FIFO, PQ and WFQ.

### A. FIFO

In this the packet first put in the queue will be first transmitted. As the router buffer space is of finite size, so when the buffer will full the packet will be discarded. This mechanism doesnot prioritize the packet transmission in a network. This queuing mechanism is not suitable for real time applications.

### B. PQ

This is a variation of FIFO queuing discipline and it marks the packet with the priority indicating in ToS field. In this queuing discipline multiple FIFO queues are implemented at router with a priority class. The packets are sorted in the buffer. The highest priority packet will be transmitted first.It is best suited for real time applications like VoIP. As higher priority packet will get transmitted first so normal packet will go to long wait in buffer and starvation situation arises.

### C. WFQ

WFQ assigns weight to each flow. The ToS field is used to identify the weight. It is a combination of PQ and FQ. In this queuing discipline packets are weighted so that lower bandwidth packet gets higher priority. WFQ supports real time interactive applications by putting traffic in front of queue with higher weight and lower bandwidth packet transmitted in smaller time. Some time it is inefficient as it increases congestion state in a network.

### V. PERFORMANCE

Performance of network is used to measures of service quality for transferring data from source to destination. There are some parameters that classify the performance of network as shown in figure 2:

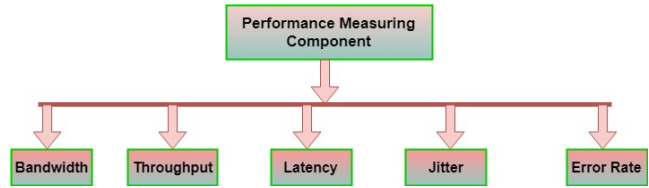

**Figure 2. Performance parameters**

### A. Bandwidth

It is the capacity to transmit the amount of data from one point to another in a given amount of time as shown in figure 3. The bandwidth measure data transfer rate. Bandwidth is usually measures in bites per second and bytes per seconds for digital communication[12].

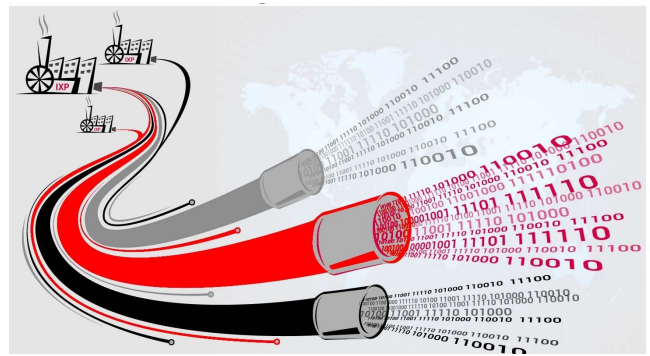

### B. Throughput

Throughput is amount of data that will be transmitted from one point to other in a given amount of time as shown in figure 4.. It is used to measure the network performance[13]. It may also define as the successful transmission of a message over a communication channel.

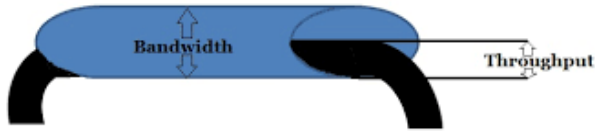

Figure 4. Throughput[13]

### C. Latency

Latency or delay is the situation arises during processing or transmitting data over a network as shown in figure 5. The network latency occurs due to delay in data communication process or transmission[14]. The latency effects the network performance.

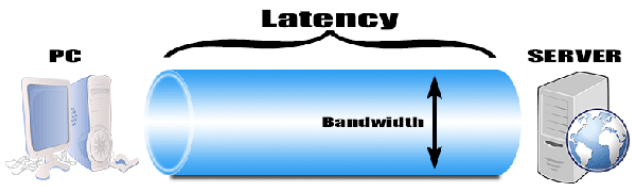

Figure 5. Latency[14]

### D. Jitter

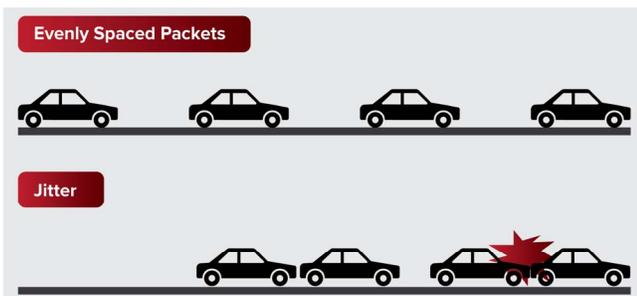

Jitter is a condition that occurs due to variation in packet arrival time s shown in figure 6. It is another kind of packet delay that causes congestion and packet loss. It effects the performance of real time application by disrupting packets[15].

Figure 6. Jitter[15]

### E. Error Rate

It is the key parameter for measuring the performance of a network. It measures number of errors encountered due to noise during data transmission over a connected network as shown in figure 7. If the **error rate** of any network is high than the network performance will be degraded and it will become less reliable for packet transmission[16].

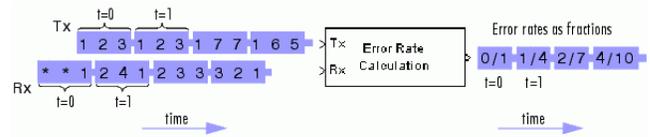

Figure 7. Error rate[16]

## VI. SIMULATION RESULTS

### A. Ethernet Delay

In case of FIFO, there is only one queue having size of 500 packets and as FIFO processed the packet on first come first serve basis and there is no priority for individual type of incoming traffic. When more packets arrive, then they get stored in a waiting queue and if queue becomes full the incoming packets will be discarded.

In WFQ packets dropped occur due to congestion in network state. In PQ the priority is higher for voice packets than WFQ and FIFO as shown in figure 8.

In case of PQ, Voice packets are assigned highest priority. After voice, video packets get priority than normal packets. Voice traffic has supreme priority so they are transmitted as soon as received, and video and normal traffic has to wait when there is incoming voice packets. Again video traffic has priority over FTP in PQ.

As WFQ handles shared buffer and when buffer gets overloaded congestion situation arises, now interface enforced queue to be limited. When congestion occurs voice packet has to wait as video packets causing packet loss as queue becomes full which is opposed to PQ where voice traffic has higher priority and need not to be waiting. So overall observation is FIFO has higher packet dropping and WFQ has semi lower and PQ is lower.

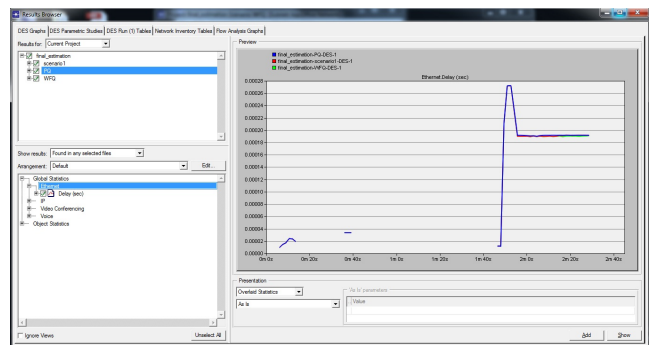

Figure 8. Packets Dropped

### B. Voice Traffic Received

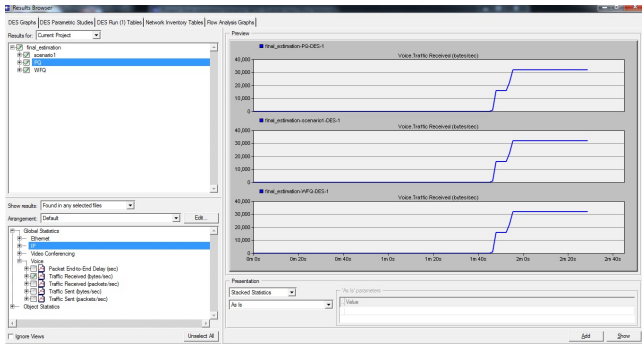

### D. Voice Packet End To End Delay

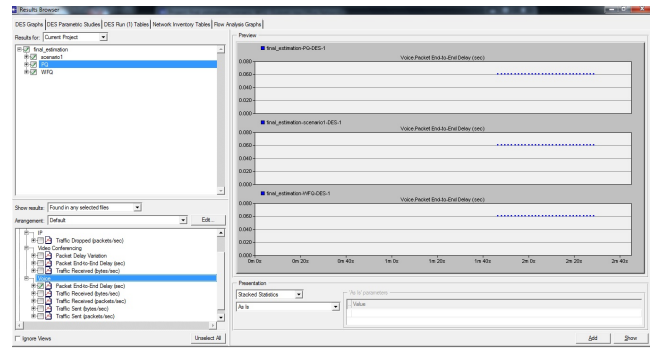

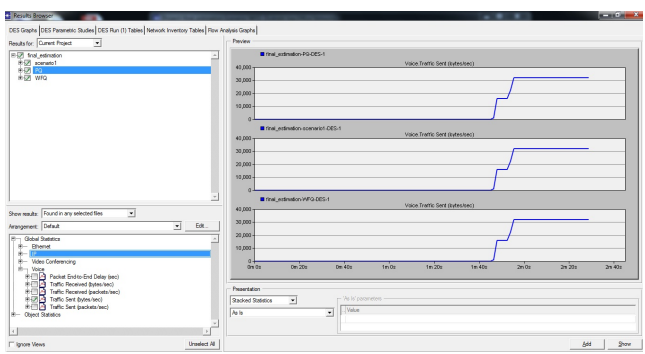

**Figure 9.(a) (b) Voice Traffic Received**

The packet has been sent through three different queuing disciplines as shown in figure 9(a) and (b) and it shows that FIFO received less voice packets than PQ and WFQ and higher packet received under PQ and WFQ.

### C. Voice Traffic Sent

**Figure 10. Voice traffic sent**

In figure 10 the traffic sent for all queuing disciplines are almost the same as there is no packet dropped that means that the network is working efficiently.

**Figure 11. Voice packet end to end delay**

Figure 11 shows that the packet transmitted from source to destination and it was found that E-E delay is higher in FIFO and lower in PQ and WFQ.As real time applications require lower delay so PQ and WFQ is better queuing policy to be used for VoIP.

### VII. CONCLUDING REMARKS

For real time communications, like VoIP the PQ is best as it allows high traffic received, than WFQ and FIFO.PQ also decreases End to End delay. FIFO is not best for real time applications as it has higher packet loss. In the future work we will study the real time applications parameters like resolution, speech quality to justify which queuing discipline may be preferred for best high packet sent and received with lower E-E delay.